\documentstyle[twocolumn,aps,amssymb,graphicx]{revtex}

\begin{document}
\draft
\title{Small world effect in an epidemiological model}

\author{Guillermo Abramson$^{1,2}$
\thanks{E-mail address: abramson@cab.cnea.gov.ar}
and Marcelo Kuperman$^1$
\thanks{E-mail address: kuperman@cab.cnea.gov.ar}}
\address{$^1$Centro At{\'o}mico Bariloche and Instituto Balseiro, 8400
S. C. de Bariloche, Argentina \\
$^2$Consejo Nacional de Investigaciones Cient{\'\i}ficas y T{\'e}cnicas,
Argentina}

\maketitle

\begin{abstract}
{\normalsize A model for the spread of an infection is analyzed
for different population structures. The interactions within the
population are described by small world networks, ranging from
ordered lattices to random graphs. For the more ordered systems,
there is a fluctuating endemic state of low infection. At a
finite value of the disorder of the network, we find a transition
to self-sustained oscillations in the size of the infected
subpopulation.}
\end{abstract}


\section{Introduction}

How does the dynamics of an infectious disease depend on the
\emph{structure} of a population? A great amount of work has been
done on the phenomenological description of particular epidemic
situations \cite{sciam,sciam2,sciam3,sciam4}. A classical
mathematical approach to these problems deals with well mixed
populations, where the subpopulations involved (typically
susceptible, infected and removed) interact in proportion to
their sizes. With these zero dimensional models it has been
possible to study, among other epidemic features, the existence
of threshold values for the spread of an infection \cite{and},
the asymptotic solution for the density of infected people
\cite{murr,bai,hop}, the effect of stochastic fluctuations on the
modulation of an epidemic situation \cite{rusos}. A second
classical approach describes spatially extended subpopulations.
In this, the geographic spread of an epidemic can be analyzed as
a reaction-diffusion process \cite{murr2,kal,kup1}.

Real populations rarely fall into either of these categories,
being neither well mixed nor lattices. Recently introduced by
Watts and Strogatz \cite{watts}, small world networks attempt to
translate, into an abstract model, the complex topology of social
interactions. Small worlds may play an important role in the
study of the influence of the network structure upon the dynamics
of many social processes, such as disease spreading, formation of
public opinion, distribution of wealth, transmission of cultural
traits, etc. \cite{watts2}. In relation to epidemiological models,
it has been shown that small world networks present a much faster
epidemic propagation than reaction-diffusion models or discrete
models based on regular lattices of a social network \cite{swep}.

In the original model of small worlds a single parameter $p$,
running from 0 to 1, characterizes the degree of disorder of the
network, respectively ranging from a regular lattice to a
completely random graph. It has been shown that geometrical
properties, as well as certain statistical mechanics properties,
show a transition at $p_c=0$ in the limit of large systems,
$N\rightarrow \infty$ \cite{barrat}. That is, any finite value of
the disorder induces the small world behavior. In this article we
show that a sharp transition in the behavior of an infection
dynamics exists at a finite value of $p$.

\section{Epidemic model}

We analyze a simple model of the spread of an infectious disease.
We want, mainly, to point to the role played by the network
structure on the temporal dynamics of the epidemic. The disease
has three stages: susceptible ($S$), infected ($I$), and
refractory ($R$). An element of the population is described by a
single dynamical variable adopting one of these three values.
Susceptible elements can pass to the infected state through
contagion by an infected one. Infected elements pass to the
refractory state after an infection time $\tau_I$. Refractory
elements return to the susceptible state after a recovery time
$\tau_R$. This kind of system is usually called $SIRS$, for the
cycle that a single element goes over. The contagion is possible
only during the $S$ phase, and only by an $I$ element. During the
$R$ phase, the elements are immune and do not infect.

The interactions between the elements of the population are
described by a small world network. The links represent the
contact between subjects, and infection can only proceed through
them. As in Watts and Strogatz model, the small worlds we study
are random networks built upon a topological ring with $N$
vertices and coordination number $2K$. Each link connecting a
vertex to a neighbor in the clockwise sense is then rewired at
random, with probability $p$, to any vertex of the system. With
probability $(1-p)$ the original link is preserved.
Self-connections and multiple connections are prohibited. With
this procedure, we have a regular lattice at $p=0$, and
progressively random graphs for $p>0$. The long range links that
appear at any $p>0$ trigger the small world phenomenon. At $p=1$
all the links have been rewired, and the result is similar to
(though not exactly) a completely random network. This algorithm
should be used with caution, since it can produce disconnected
graphs. We have used only connected ones for our analysis.

Time proceeds by discrete steps. Each element is characterized by
a time counter $\tau_i(t)=0,1,\dots,\tau_I+\tau_R\equiv\tau_0$,
describing its phase in the cycle of the disease. The
epidemiological state $\pi_i$ of the element ($S$, $I$ or $R$)
depends on this phase in the following way:
\begin{equation}
\begin{array}{ll}
\pi_i(t)=S & \;\;\mbox{if}~\tau_i(t) = 0 \\
\pi_i(t)=I & \;\;\mbox{if}~\tau_i(t) \in (1,\tau_I) \\
\pi_i(t)=R & \;\;\mbox{if}~\tau_i(t) \in (\tau_I\!+\!1,\tau_0)
\end{array}
\end{equation}

The state of an element in the next step depends on its current
phase in the cycle, and the state of its neighbors in the
network. The rules of evolution are the following:
\begin{equation}
\begin{array}{ll}
\tau_i(t\!+\!1)=0 & \mbox{if $\tau_i(t)=0$ and no infection occurs} \\
\tau_i(t\!+\!1)=1 & \mbox{if $\tau_i(t)=0$ and $i$ becomes infected} \\
\tau_i(t\!+\!1)=\tau_i(t)\!+\!1 & \mbox{if}\; 1\leq\tau_i(t)<\tau_0  \\
\tau_i(t\!+\!1)=0 & \mbox{if}\; \tau_i(t)=\tau_R
\end{array}
\end{equation}
That is, a susceptible element stays as such, at $\tau=0$, until
it becomes infected. Once infected, it goes (deterministically)
over a cycle that lasts $\tau_0$ time steps. During the first
$\tau_I$ time steps, it is infected and can potentially transmit
the disease to a susceptible neighbor. During the last $\tau_R$
time steps of the cycle, it remains in state $R$, immune but not
contagious. After the cycle is complete, it returns to the
susceptible state.

The contagion of a susceptible element by an infected one, and the
subsequent excitation of the disease cycle in the new infected,
occur stochastically at a local level. Say that the element $i$
is susceptible, and that it has $k_i$ neighbors, of which
$k_{\text{inf}}$ are infected. Then, $i$ will become infected
with probability $k_{\text{inf}}/k_i$. Observe that $i$ will
become infected with probability 1 if all its neighbors are
infected. There are other reasonable choices for this mechanism.
For example, if the susceptible had a probability $q$ of
contagion with each infected neighbor, we would have a
probability of infection $[1-(1-q)^{k_{\text{inf}}}]$. We have
tested that both these criteria give qualitatively the same
results for values of $q\lesssim 0.2$.

\section{Numerical results}

We have performed extensive numerical simulations of the
described model. Networks with $N=10^3$ to $10^6$ vertices have
been explored, with $K=3$ to $10$. A typical realization starts
with the generation of the random network and the initialization
of the state of the elements. An initial fraction of 0.1
infected, and the rest susceptible, was used in all the results
shown here. Other initial conditions have been explored as well,
and no changes have been observed in the behavior. After a
transient a stationary state is achieved, and the computations
are followed for several thousand time steps to perform
statistical averages.

\begin{figure}[tbp]
  \centering
  \resizebox{\columnwidth}{!}{\includegraphics{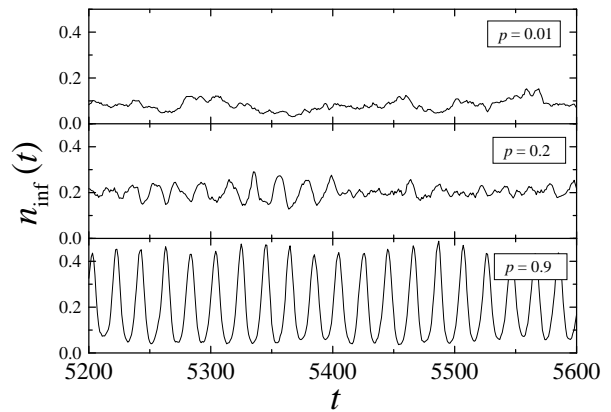}}
\caption{Fraction of infected elements as a function of time. Three time
series are shown, corresponding to different values of the
disorder
parameter $p$, as shown in the legends. Other parameters are: $N=10^4$, $K=3$%
, $\tau_I=4$, $\tau_R=9$, $N_{\text{inf}}(0)=0.1$.}
\label{tempo}
\end{figure}

We show in Fig. \ref{tempo} part of three time series displaying
the fraction of infected elements in the system,
$n_{\text{inf}}(t)$. The three curves correspond to systems with
different values of the disorder parameter: $p=0.01$ (top), 0.2
(middle) and 0.9 (bottom). The three systems have $N=10^4$ and
$K=3$, and infection cycles with $\tau_I=4$ and $\tau_R=9$. The
initial state is random with $n_{\text{inf}}(0)=0.1$. The 400 time
steps shown are representative of the stationary state. We can
see clearly a transition from an endemic situation to an
oscillatory one. At $p=0.01$ (top), where the network is nearly a
regular lattice, the stationary state is a fixed point, with
fluctuations. The situation corresponds to that of an endemic
infection, with a low and persistent fraction of infected
individuals. At high values of $p$---like the case with $p=0.9$
shown in the figure (bottom)---large amplitude, self-sustained
oscillations develop. The situation is almost periodic, with a
very well defined period and small fluctuations in amplitude. The
period is slightly longer than $\tau_0$, since it includes the
average time that a susceptible individual remains at state $S$,
before being infected. Epidemiologically, the situation resembles
the periodic epidemic patterns typical of large populations
\cite{sciam3}. A mean field model of the system, expected to
resemble the behavior at $p=1$, can be easily shown to exhibit
these oscillations. The transition between both behaviors is
apparent---in this relatively small system---at the intermediate
value of disorder $p=0.2$, shown in the middle curve. Here a low
amplitude periodic pattern can be seen, appearing and
disappearing again in a background of strong fluctuations.
Moreover, the mean value of infection is seen to grow with $p$.

The formation of persistent oscillations corresponds to a
spontaneous synchronization of a significant fraction of the
elements in the system. Their phases $\tau_i(t)$ in the epidemic
cycle become synchronized, and they go over the disease process
together, becoming ill at the same time, and recovering at the
same time. We can characterize this behavior with a
synchronization parameter \cite{kura}, defined as:
\begin{equation}
\sigma(t)=\left|\frac{1}{N}\sum_{j=1}^N e^{i\,\phi_j(t)}\right|,
\label{sigma}
\end{equation}
where $\phi_j=2\pi (\tau_j-1) /\tau_0$ is a geometrical phase
corresponding to $\tau_j$. We have chosen to let the states
$\tau=0$ out of the sum in (\ref{sigma}), and take into account
only the deterministic part of the cycles.

When the system is not synchronized, the phases are widely spread
in the cycle and the complex numbers $e^{i\phi}$ are
correspondingly spread in the unit circle. In this situation
$\sigma$ is small. On the other hand, when a significant part of
the elements are synchronized in the cycle, $\sigma$ is large. The
synchronization would be strictly $\sigma=1$ if all the elements
were at the same state at the same time. However, such a state
would end up in $N_{\text{inf}}=0$ after $\tau_0$ time steps, and
the epidemic would end since the system is closed and no
spontaneous infection of susceptibles is being taken into account
in the model.

\begin{figure}[tbp]
  \centering
  \resizebox{\columnwidth}{!}{\includegraphics{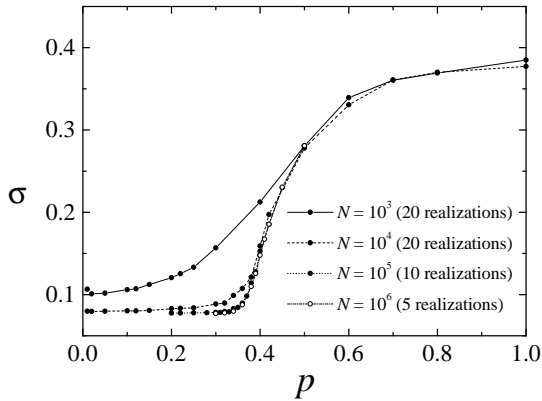}}
\caption{Synchronization of the system as a function of the disorder
parameter $p$. Three curves are shown, corresponding to different
system sizes $N=10^3$, $N=10^4$, $N=10^5$ and $10^6$, as shown in
the legend. Each point corresponds to a time average of 2000 time
steps, and a subsequent average over a number of realizations of
the networks and the initial
condition, as shown in the legend. Other parameters are: $K=3$, $\tau_I=4$, $%
\tau_R=9$, $N_{\text{inf}}(0)=0.1$.}
\label{sync}
\end{figure}

In Fig. \ref{sync} we show the synchronization parameter
$\sigma$, obtained as a time average of $\sigma(t)$ over 2000
time steps and a subsequent average over realizations of the
system. Several curves are shown, corresponding to system sizes
$N=10^3, 10^4, 10^5$ and $10^6$. A transition in the
synchronization can be seen as $p$ runs from 0 to 1. The
transition becomes sharper for large systems, at a value of the
disorder parameter $p_c\approx0.4$. It is worthwhile to remark
that the transition to synchronization occurs as a function of
the structure of the network, contrasting the phenomenon of
synchronization as a function of the strength of the interaction,
as in other systems of coupled oscillators \cite{heagy}. This
behavior is observed for a wide range of values of $\tau_I$ and
$\tau_R$. The amplest oscillations take place around
$\tau_I/\tau_R=1$. They disappear when $\tau_I$ is significantly
greater or smaller than $\tau_R$.

\begin{figure}[tbp]
  \centering
  \resizebox{\columnwidth}{!}{\includegraphics{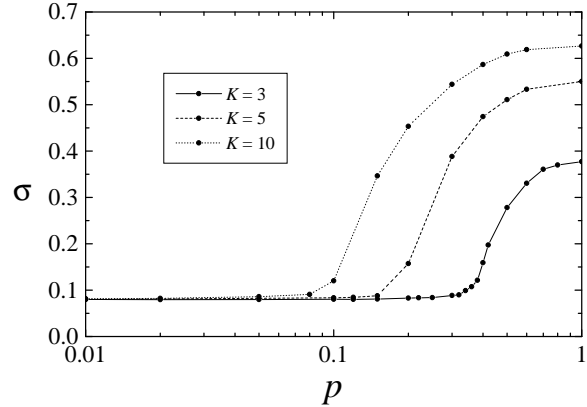}}
\caption{Synchronization of the system as a function of the disorder
parameter $p$. Three systems with different values of the
coordination
number $K$ are shown. All of them have $N=10^4$, $\tau_I=4$, $\tau_R=9$, $N_{%
\text{inf}}(0)=0.1$.}
\label{difk}
\end{figure}

All the systems shown in figures \ref{tempo} and \ref{sync} have
$K=3$. We have explored higher values of $K$ as well. The picture
is qualitatively the same, with a sharp transition from a
quasi-fixed point to a quasi-limit cycle at a finite value of
$p$. The critical value $p_c$ shifts toward lower values for
growing $K$. This is reasonable, since higher values of $K$
approaches the system to a globally coupled one (at $K=N$ all the
elements interact with every other one, even at $p=0$). So, the
mean field behavior (the oscillations) can be expected to occur
at lower values of $p$. In figure \ref{difk} this effect can be
seen for a system with $N=10^4$ and growing values of $K=3$, 5
and 10.

\section{Discussion}

Why does this transition to synchronization take place?
Unfortunately we do not have analytical arguments to describe
this yet. As mentioned before, a mean field model can be shown to
predict the oscillations, but this can only be expected to
describe the network at $p=1$, and it does not shed light on the
nature of the transition at lower values of disorder. We can only
advance some conjectures, based in our observation of the
dynamical behavior of the system in real time on the computer
screen.

An explanation involving $L(p)$, the typical distance between
pairs of elements (defined as the size of the minimum path
connecting two elements) is not plausible, since $L$ is known to
behave critically with $p_{c}=0$, and we observe the transition
at $p>0$. Complementarily to $L$, small world networks can be
described by the degree of clusterization $C(p)$ \cite{cluster}.
At low values of $p$, the networks are rather regular and highly
clustered. As $p$ approaches 1, $C$ decreases. The crossover from
high to low clusterization occurs at a higher value of $p$,
compared to that observed in the decay of $L$. For systems with
$K=3$, such as those mainly studied here, we have that this
crossover is mostly concentrated between $ p=0.1$ and $p=0.5$,
precisely where the onset of oscillations occurs. Moreover, the
change in the average clusterization $C(p)$ is accompanied by a
corresponding one in the distribution of the clusterization at
the element level, $c_{i}(p)$. Highly ordered networks, at low
$p$, show not only a large value of the average clusterization
$C(p)$ but also a small dispersion around it. On the other
extreme, highly disordered networks, with $p>0.5$, exhibit a low
average clusterization and also a low dispersion around it. There
is an intermediate range of $p$, between 0.1 and 0.5, where the
average clusterization shifts from high to low, but the
distributions are wide. This indicates that, in this range, the
system is a mixture of highly ordered, highly clustered regions
and random, lowly clustered ones. If we consider that the
clustering structure determines a partition of the whole system
into smaller, interacting, subsystems, the global behavior of the
system could be interpreted as a superposition of the subsystems'
dynamics. When $p$ is small the existence of large clusters
(essentially one-dimensional in our model) inhibits the
oscillatory behavior because, once the infection breaks into such
a region, it remains restricted to it a long time until all the
individuals have completed the cycle, orderly and
deterministically. As $p$ increases, the number of big, ordered
regions decreases. Elements within small regions with some long
range links go through the infection cycle and become infected
again before long. Some degree of synchronization can be seen
here (Fig. \ref{tempo}, center, corresponds to $p=0.2$). When a
critical value of $p$ is reached, and the system is essentially
composed of enough small regions of low clusterization and a
similar local dynamics, the synchronized, periodic global
behavior establishes spontaneously.

In summary, we have observed a transition at a finite value of
the disorder in a small world model. The dynamical behavior of an
SIRS epidemiological model changes from an irregular,
low-amplitude evolution at small $p$ to a spontaneous state of
wide amplitude oscillations at large $p$. This may be related to
observed patterns in real epidemics \cite{sciam3}, where an
effect of the social structure is observed in the dynamics of the
disease.

The authors thank D. H. Zanette for fruitful discussions.


\begin{references}
\bibitem{cluster} $C(p)$ is an average, over the system, of the
local clusterization $c_i(p)$. This is defined as the number of
neighbors, of element $i$, that are neighbors among themselves,
normalized to the value that this would have if all of them were
connected to one another---namely $k_i(k_i-1)/2$.
\end{references}
\end{document}